# Large Conductance Variations in a Mechanosensitive Single–Molecule Junction

*Davide Stefani,[1,‡] Kevin J. Weiland,[2,‡] Maxim Skripnik,[3,4,‡] Chunwei Hsu,[1] Mickael L. Perrin,[1,5] Marcel Mayor,[2,6,7]\* Fabian Pauly,[3,4]\* and Herre S. J. van der Zant[1]\**

[1]Kavli Institute of Nanoscience, Delft University of Technology, 2600 GA Delft, The Netherlands

[2]Department of Chemistry, University of Basel, 4056 Basel, Switzerland

[3]Okinawa Institute of Science and Technology Graduate University, Onna-son, Okinawa 904-0395, Japan

[4]Department of Physics, University of Konstanz, 78457 Konstanz, Germany

[5]Empa, Swiss Federal Laboratories for Materials Science and Technology, Transport at Nanoscale Interfaces Laboratory, 8600 Dübendorf, Switzerland

[6]Karlsruhe Institute of Technology (KIT), P.O. Box 3640, 76021 Karlsruhe, Germany

[7]Lehn Institute of Functional Materials, School of Chemistry, Sun Yat-Sen University, Guangzhou 510275, China

[‡]These authors contributed equally to this work.

**Abstract:** The appealing feature of molecular electronics is the possibility of inducing changes in the orbital structure through external stimuli. This can provide functionality on the single molecule level, which can be employed for sensing or switching purposes, if the associated conductance changes are sizable upon application of the stimuli. Here, we show that the room-temperature conductance of a spring-like molecule can be mechanically controlled up to an order of magnitude by compressing or elongating it. Quantum chemistry calculations indicate that the large conductance variations are the result of destructive interference effects between the frontier orbitals that can be lifted by applying either compressive or tensile strain to the molecule. When periodically modulating the electrode separation, a conductance modulation at double the driving frequency is observed, providing a direct proof for the presence of quantum interference. Furthermore, oscillations in the conductance occur when the stress built up in the molecule is high enough to allow the anchoring groups to move along the surface in a stick-slip-like fashion. The mechanical control of quantum interference effects results in the largest gauge factor reported for single-molecule devices up to now, which may open the door for applications in, *e.g.*, a nano-scale mechanosensitive sensing device functional at room temperature.

In recent years, studies on single-molecule junctions have rapidly become a mature research field.[1,2] The combination of the structural diversity accessible by synthetic chemistry with the continuously improving skills of experimental and theoretical physics enabled the exploration of individual molecules as the tiniest functional building blocks for electronic circuits and sensors.[3,4] The steady refinement of the break-junction technique also allowed to study correlations between mechanical manipulations and transport features at the single-molecule level in a systematic way.[5] Some examples are binary switching through mechanical control of the metal–molecule contact geometry[6] or stereoelectronic effects,[7] mechanical stress sensitive redox chromophores[8] or coordination compounds that show spin-state switching under mechanical stress.[9] Furthermore, the sensitivity of the technique makes it possible to observe intermolecular behaviours like π-stacking,[10,11] and more recently, the interdependence of conductance and frontier orbitals involved in the π-stacking[12,13] could be directly demonstrated.[14]

Particularly interesting are destructive quantum interference effects leading to a strong suppression of electron transmission at specific energies, which make them an ideal feature for applications in, *e.g.*, thermo-[15] or voltage-dependent switching.[16] Their manipulation has been reported by external means including solid state[17] or electrochemical gating,[18] humidity,[19] and sliding of π-stacked molecules relatively to each other.[14] Deliberate manipulation of the latter, however, remains elusive as it requires strict temperature conditions and is based on intermolecular interactions. In particular, the intermolecular character requires the coincidental presence of two molecules inside the junction. For this reason, approaches that intramolecularly imitate intermolecular π-stacking move into the focus of interest. Along these lines, the [2.2]paracyclophane (PC) compound is highly appealing.[20] First described by Farthing *et al.* in 1949, it consists of two stacked benzene rings which are mechanically stabilised by two non-conjugated linkers.[21] Integrated as central unit of an oligo-phenylene-ethynylene (OPE) rod with terminal binding groups to gold electrodes (Fig. 1), we show here that using a mechanically controlled break junction (MCBJ) the π-stacking (and therefore the conductance) can be modulated by exerting a mechanical shear force to it. Simulations based on density functional theory (DFT) reveal a sensitive correlation between electrode displacement and molecular conductance, which is interpreted in terms of quantum interference effects between the frontier orbitals.

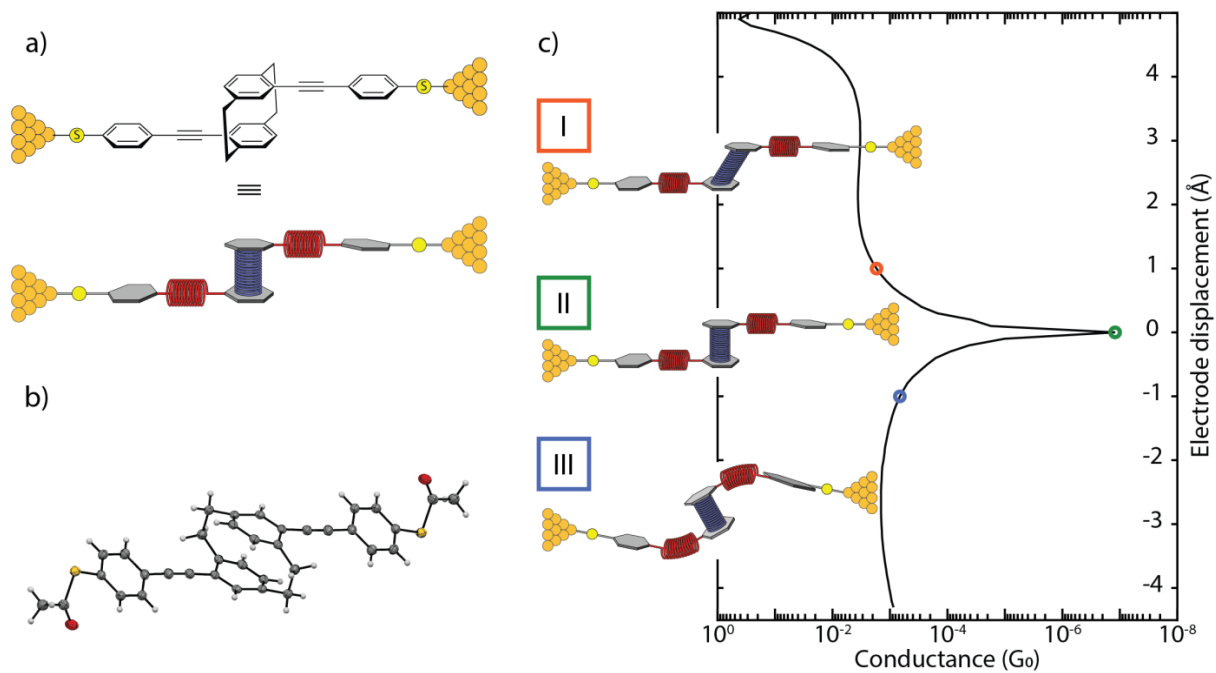

**Figure 1.** a) Schematic illustration of the break junction experiment of the OPE-linked PC molecule trapped between nanoelectrodes (top panel), together with a presentation displaying, as springs, the combinations of mechanosensitive structures in the molecular design (bottom drawing in this panel). b) Solid state molecular structure determined by single crystal X-ray analysis. c) Possible behaviour of the molecule under applied force: I) elongation of the molecule under pulling force of the electrodes; II) junction with the molecule in its relaxed configuration; III) compression results in a shortening of the overall junction length. The simulated conductance (in units of the conductance quantum $G_0 = 2e^2/h$) as a function of the applied mechanical stress is displayed as the drawn line; the three cases I-III are indicated by the coloured circles.

**Results**

The employed molecule and the conceptual idea behind the mechanical conductance manipulation are displayed in Fig. 1. The molecule comprises a motif where two ethynylphenylthiols are bound to PC in such a way that the connection resembles a *para*-substitution in benzene.[22,23] The thiol groups are connected in the *para* position of the outer benzene rings with respect to the PC building block, directing the current path through the PC, and offering considerable mechanical stability between gold electrodes.[24] The molecular motif and substitution pattern allow for flexibility by stretching of the PC; the ethynylphenylthiol building block, on the other hand, only offers limited movement upon application of a pulling force on the molecule (Fig. 1c). Precedence for mechanical flexibility of PC derivatives exists, as high-pressure solid state absorption experiments show shortened distances of its benzene rings.[25] When the thiol groups are moved apart, the applied tensile force is relayed to the central PC unit along the axis connecting the anchoring groups (Fig. S4). Our DFT calculations show that the alkynes are more susceptible to compressive motions, whereas the PC subunit only stretches after these are fully extended. It is noteworthy that electrical transport studies through monolayers consisting of the proposed molecular rods and similar PC-containing subunits have been reported[23,26]. However, the limited control over the number of molecules inside the junction made the interpretation of the results challenging.

**Single-molecule experiments**

The target molecule was synthesised by adapting literature-known procedures;[27,28] its structure was unambiguously verified by single-crystal X-ray analysis (Fig. 1b). Details of the synthetic protocols are provided in the Supporting Information (SI) together with the analytical data corroborating its identity, which is in agreement with the data already reported.[23] The molecular conductance was investigated using the MCBJ technique under ambient conditions. In this technique, atomically sharp electrodes are formed in a lithographically defined gold wire and repeatedly opened and closed with sub-nanometre accuracy. The measurement of the conductance as the electrodes are continuously moved further apart constitutes a so-called breaking trace. Further details about the MCBJ setup and the measuring technique have been described elsewhere.[29,30]

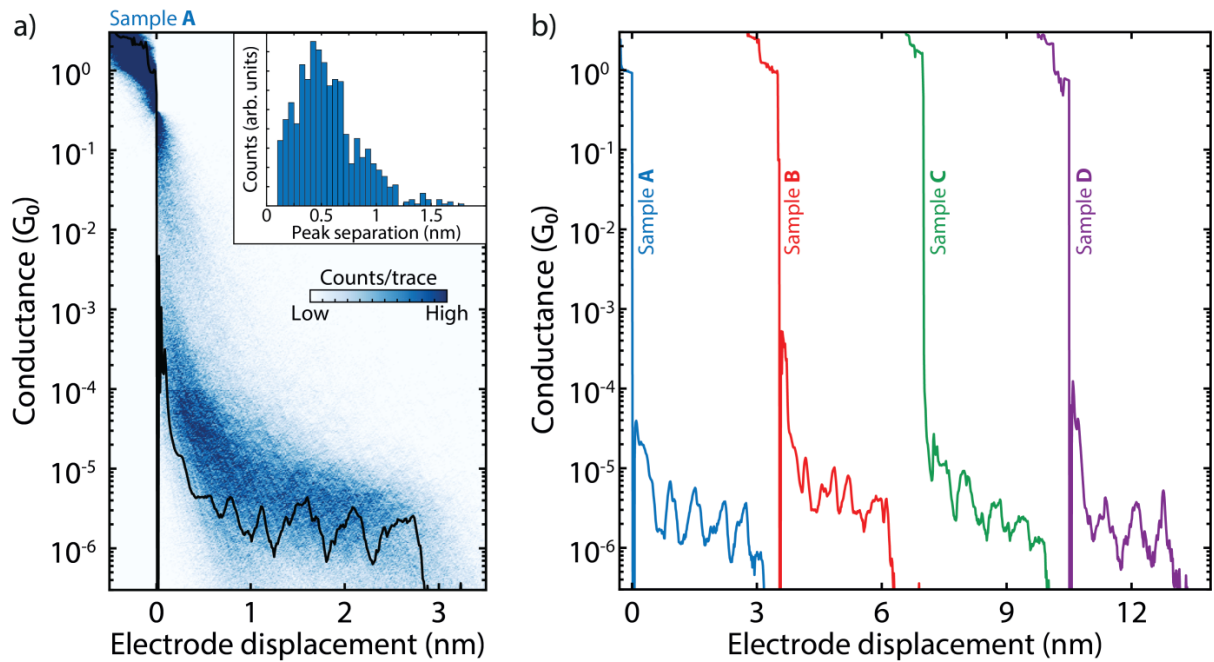

**Figure 2.** a) Two-dimensional histogram built from 3,000 consecutive breaking traces collected after deposition of the solution containing the molecule displayed in Fig. 1. The applied bias is 100 mV and the electrode speed is 4.0 nm/s. A single breaking trace (black line) has been overlaid as an example. The inset shows the peak separation distribution. b) Examples of breaking traces, showing oscillations in conductance as a function of displacement. The traces are taken from four different samples and are offset

in the x-axis for clarity. The first trace comes from sample **A**, shown in a); sample **B**, **C** and **D** are presented in the Supporting Information (see Figs. S6-S9) alongside more examples of breaking traces.

Fast-breaking measurements were performed to characterise the conductance of the OPE-linked PC molecules.[31] The two-dimensional histogram in Fig. 2a shows the distribution of conductance versus electrode displacement. Fitting a log-normal distribution to the one-dimensional histogram constructed from the same data (Fig. S5a) reveals that the most probable conductance value is $3.7 \cdot 10^{-6}$ $G_0$, where $G_0 = 2e^2/h$ is the quantum of conductance. Interestingly, inspection of individual traces shows the appearance of pronounced oscillations in the conductance as the electrodes are pulled from one another, as illustrated in Fig. 2b for four different samples. These oscillations are found in a large fraction of the molecular traces (40 %) and have an amplitude up to an order of magnitude. The inset of Fig. 2a presents a histogram of the spacing of the individual peaks, showing a periodicity of about 0.5 nm (see Supporting Section II-1 for more details). Note that such oscillations are absent in the control measurements without the molecule (Fig. S10).

To investigate in more detail the dependence of the molecular conductance on the electrode displacement, we performed conductance measurements with a modulated electrode spacing.[32] In this experiment, the MCBJ was initially stretched to a few-atom width (about 3 $G_0$) and allowed to self-break by its surface tension to form atomically sharp electrodes.[33] Then, the electrodes were separated by a distance equal to the length of the molecule to recognise, by evaluating the conductance, whether the trapping of a molecule occurred: if the conductance is found to be larger than $10^{-6}$ $G_0$, a molecule is presumably connected between the two electrodes, and a triangular wave is applied to the piezoelectric stack that controls the electrode positions (Fig. 3a).

Note, that a higher voltage on the piezoelectric stack corresponds to a larger electrode separation. Hundreds of such so-called distance-modulation traces are collected and from them a conductance histogram is built, similarly to that obtained for fast-breaking measurements. Fitting a log-normal distribution to this histogram yields a peak at $2.7 \cdot 10^{-6}$ $G_0$, a value close to that found for fast-breaking measurements (Fig. S5b). Thus, molecular traces in this measurement appear at approximately the same conductance values as found for the fast-breaking measurements and have a long lifetime, consistently surviving for the entire modulation time of 120 s.

As illustrated in Fig. 3a and Figs. S12-S16, as the gap size increases the conductance can either increase or decrease. In the former case we define the conductance changes to be *in-phase* with the gap size modulation (orange curve in Fig. 3a). In the latter it is the other way around: the conductance change is in *antiphase* with the gap size modulation (blue curve in Fig. 3a). About 32% of the molecular traces show in-phase conductance variations, 28% appear to respond in antiphase, and about 40% show a mixture of both or a more complex response (Figs. S13-S17). Most of the conductance traces show these conductance variations at the same frequency as the driving modulation (Fig. S18); however, surprisingly, many (31%) respond at double the driving frequency. The green curve in Fig. 3a is an example of this.

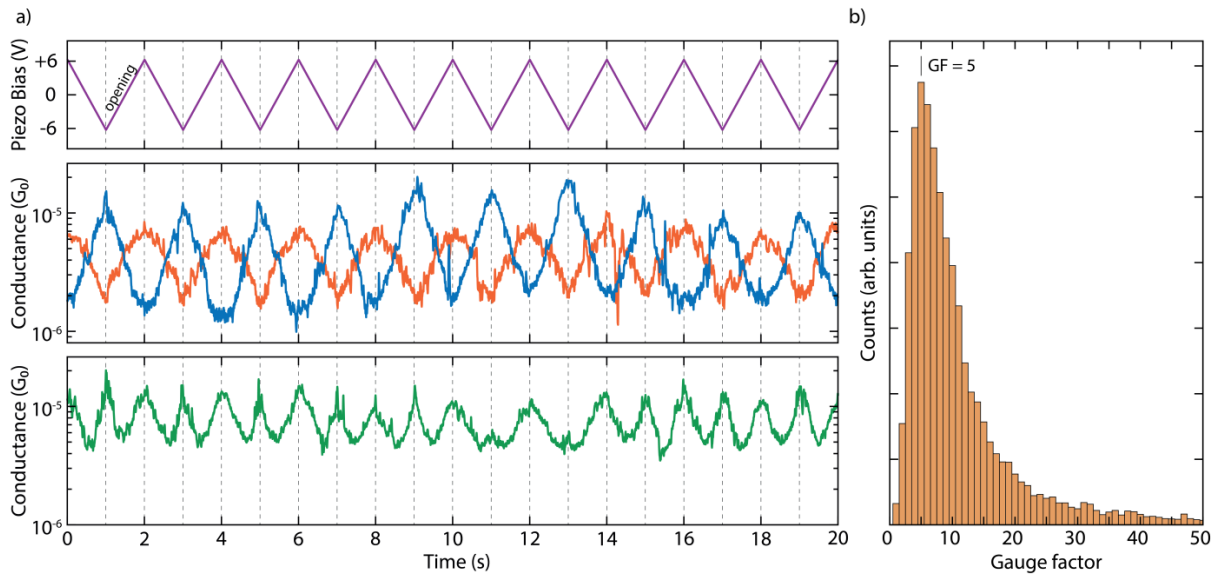

**Figure 3**. a) Examples of distance-modulation traces (sample **B**). The blue, orange and green lines (middle and bottom panel) represent three different conductance measurements, whereas the purple line (top panel) represents the voltage applied to the piezoelectric stack. The applied piezo voltage translates into a peak-to-peak distance of 5.0 Å and a positive voltage corresponds to a larger electrode distance. The total modulation time is 120 s at a frequency of 0.5 Hz. The conductance responds to the electrode-separation modulation either in-phase (orange), in antiphase (blue), or with double the modulation frequency (green). An example of a full measurement, extending over 120 s, is shown in Figs. S17 and S19. b) Distribution of absolute values of gauge factors obtained from 123 selected traces of the distance-modulation measurements performed on sample **B**. The number in the figure corresponds to the peak value of the distribution.

**Theoretical calculations**

To shed more light on these experimental observations, DFT-based calculations were carried out. For the evaluation of the conductance we used a proprietary framework.[34] To start, we place the molecule between two pyramidal gold leads, as shown in Fig. 4b. The upper electrode exhibits an atomically sharp tip, while the tip atom is removed for the lower one. By placing a terminal sulfur atom of the molecule at the hollow site of the lower pyramid, it binds to three gold atoms. Compared to the sulfur-gold bonding at the top lead, the stronger bonding of the sulfur atom to the hollow site ensures a stable mechanical connection, and the sulfur atom at the upper electrode starts sliding down the gold facets, as the contact is being stretched.

The total energy and conductance of this system show pronounced jumps at certain displacements during the gap opening. The snapshots at these displacements are shown in Fig. 4b and reveal the expected movement of the sulfur atom. The displacement between snapshots iii and v (just after the sulfur atom jumped to the next gold atom) amounts to 2.8 Å, which is close to the gold-gold binding length of 2.89 Å in the simulated leads. The molecule in snapshots i, iii and v is close to its relaxed gas phase configuration (discussed below) and exhibits a low conductance. Upon further stretching, the conductance rises until it eventually levels off and reaches a local maximum at snapshots ii and iv. At this point, the sulfur anchor slips onto the next gold atom, thus releasing the mechanical tension in the molecule and restoring the conductance to a low value (iii, v). After the sulfur atom has reached the last gold atom of the upper lead, it finally loses its connection: the junction breaks, and the molecule snaps back, as shown in snapshot vi. Thus, distancing of the electrodes leads to a stick-slip like motion of the molecule along the surface of one of the electrodes.

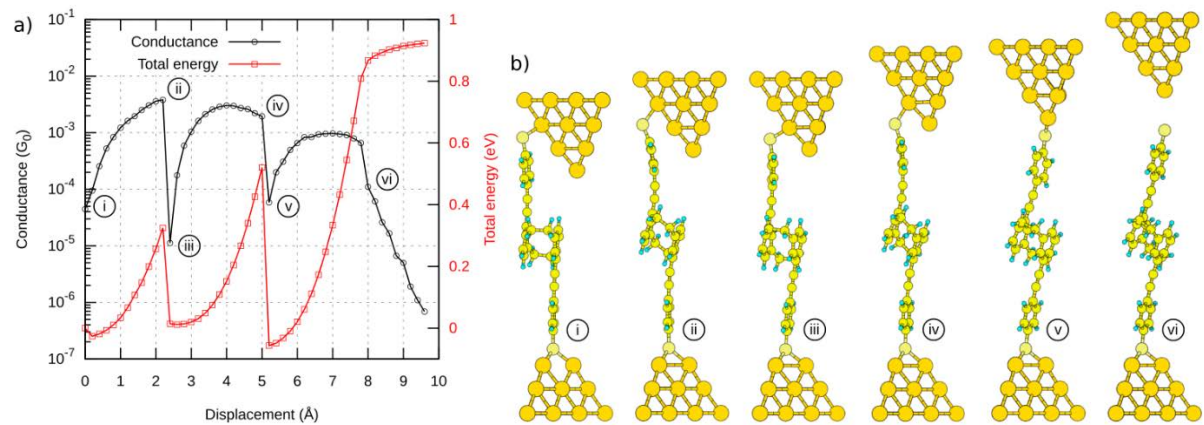

**Figure 4.** a) Calculated conductance and total energy of the system during gap opening. b) Selection of snapshots illustrating the stick-slip motion. A video of the simulated stick-slip motion can be found in the SI.

In a detailed analysis of the displacement-dependent conductance, we study the molecule between two hollow leads, where each sulfur binds to three gold atoms of the respective lead. This allows us to concentrate on the deformations of the actual molecule: the rigid bonding of the hollow-hollow configuration ensures that the lead displacement is directly passed to the molecule, minimizing deformations of the lead-molecule bonds. Starting from the configuration with minimal energy, corresponding to a molecule close to its relaxed state in the gas phase, the leads are either separated farther apart (positive displacement) or brought closer together (negative displacement), thereby stretching or compressing the molecule. The resulting conductance shows a pronounced dip at a well-determined displacement, which we define here as zero (see Fig. 5b). The conductance rapidly increases when the molecule is either stretched or compressed from this position. The deformation in the stretched molecule is mainly identified with the shifting of the stacked benzene rings.

By evaluating the transmission in an energy range between -2 eV and +3 eV around the Fermi energy $E_F$ for each displacement step, we obtain the transmission map in Fig. 5c. It reveals a transmission valley (purple diagonal line) between the traces related to the molecular frontier orbitals (yellow horizontal lines). The conductance dip in Fig. 5b can be traced back to the intersection of the transmission valley and the Fermi energy. In other words, the energy position of the dip can be tuned by the lead separation. In the following, we present the underlying mechanism based on quantum interference of the molecular frontier orbitals.

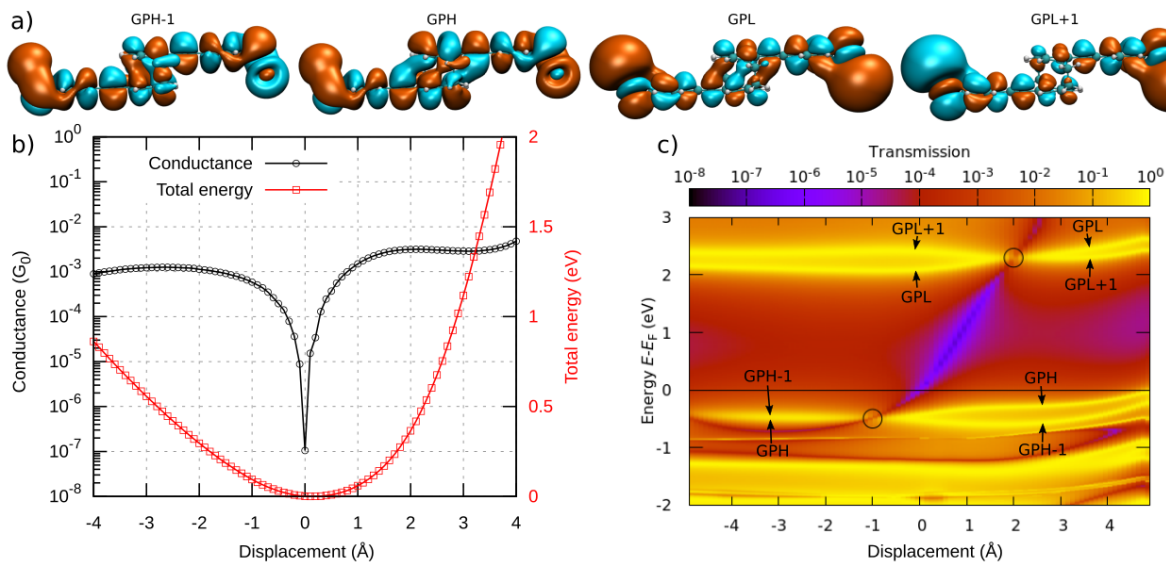

**Figure 5.** a) Frontier orbitals of the molecule. The orbitals are either symmetric (GPH, GPL) or antisymmetric (GPH-1, GPL+1) with respect to the centre of the molecule. They are shown in gas phase for clarity (with the sulfur atoms "terminated" with one gold atom each, as they do not change significantly when the gold leads are attached to the molecule). b) Conductance (horizontal black line in panel c) and total energy. c) Transmission map of the molecule between two leads from DFT calculations. The horizontal yellow traces in the map arise from orbitals which can be related to the gas phase frontier orbitals. An antiresonance is observed between the frontier orbital traces. It shifts in energy as the displacement is varied. The positions at which the pairs GPH-1, GPH and GPL, GPL+1 degenerate are marked with circles.

A closer look at Fig. 5c reveals HOMO and LUMO pairs that arise from the HOMO and LUMO states of the OPE units and are typically slightly split by the weak coupling through the PC core. To distinguish the character of the HOMO-1, HOMO, LUMO and LUMO+1, we relate them to the frontier orbitals in the gas phase. For this purpose, we introduce the abbreviations GPH-1 to GPL+1, where GPH/L denotes the gas phase HOMO/LUMO. The orbitals are either symmetric (GPH, GPL) or antisymmetric (GPH-1, GPL+1) with respect to the centre of the molecule. The crucial aspect is now that the energy of the frontier orbitals depends on the displacement. The states within HOMO and LUMO pairs, GPH-1 and GPH as well as GPL and GPL+1, eventually change their energetic order in the studied displacement window between around -4 to 4 Å. The displacements of the degeneracy points, at which these reversals take place, differ for the occupied and unoccupied states. They are located at $d = -1.0$ Å and $d = +2.0$ Å, respectively, as marked by circles in Fig. 5c. According to the theory of quantum interference[12,35] the orbital symmetry leads to a pronounced destructive interference feature in the HOMO-LUMO gap between displacements from -1.0 to 2.0 Å, when the HOMO and LUMO pairs are ordered as in the gas phase. Outside of this displacement window, the molecular orbital pairs GPH-1, GPH or GPL, GPL+1 rearrange in energy, thereby lifting the condition for destructive interference. Indeed, we can reproduce the main features of the conductance map by considering the displacement-dependent energies of the four frontier orbitals and their symmetries, as corroborated further in the SI (Section III).

**Discussion and concluding remarks**

With the insights provided by the DFT calculations, the pronounced conductance oscillations can be explained through quantum interference of frontier orbitals in combination with the molecule acting as a spring when subject to mechanical deformations. Its relaxed conformation corresponds to the situation in which the antiresonance originating from destructive interference of the HOMO and LUMO is in the vicinity of the Fermi energy, yielding a low-conductance state. By stretching or compressing the molecule, the antiresonance is moved away from the Fermi energy, which leads to an increase of the conductance. The experimentally observed oscillations in conductance during continuous opening of the junction (Fig. 2b) can then be associated to the stick-slip motion of the anchoring sulfur atoms on the gold surface, as this process releases the built-up mechanical strain in a semi-periodic fashion. Ab-initio molecular dynamics calculations at room temperature predict that Au-Au bonds should break instead of the Au-S bonds.[36] However, if a gold adatom attached to a sulfur anchor was dragged along the gold electrode instead of the S itself, this would not lead to a qualitative change of the stick-slip picture.

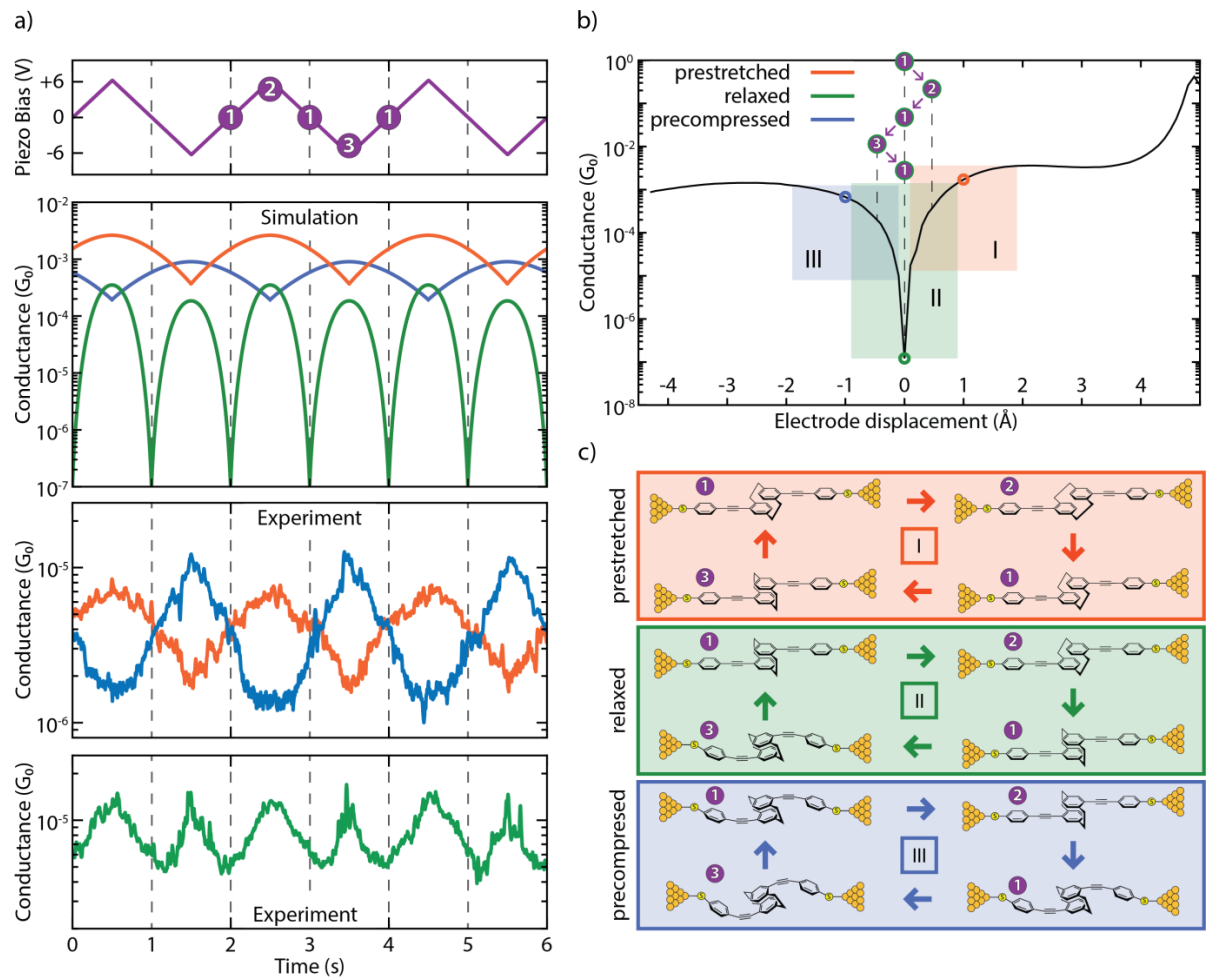

**Figure 6.** a, top panels) Simulation of conductance traces (second panel) when the electrode displacement is periodically modulated (top panel) for three different trapping configurations: prestretched (orange), relaxed (green) and precompressed (blue). Centres of oscillation at +1, 0 and -1 Å respectively; amplitude: 0.5 Å. a, bottom panels) Zoom-in of the experimental distance-modulation traces presented in Fig. 3. b) Calculated conductance vs. electrode displacement. The blue, green and orange areas (I, II and III, respectively) show the portion of the curve spanned in the case of different starting positions (precompressed, relaxed and prestretched, represented by circles in the same colours). The numbers in purple circles represent the position of the electrodes along the oscillation period in the case of a relaxed trapping configuration. c) Schematics of the molecular configurations along a period of electrode distance-modulation. Different starting configurations are represented with different colours: prestretched in orange, relaxed in green and precompressed in blue. White numbers in purple circles represent the position of the electrodes along the oscillation period. Note, that the molecular compression/elongation in the simulation is 0.5 Å, a value smaller than the gap size variation in the experiment. This can be rationalised by the elastic response of the sulfur-gold connections and of the gold atoms in the electrodes themselves.

The different behaviours (in-phase, antiphase, frequency doubling) observed in the distance-modulation measurements (simulation: second panel in Fig. 6a; experiment: bottom two panels in Fig. 6a) can be attributed to variations in the initial molecular configuration at the beginning of the modulation. Traces which are in phase with the gap modulation are related to molecules which are prestreched in the starting configuration (orange panel in Fig. 6c and "Prestretched" video in SI). The starting point is at positive displacement and the oscillation takes place at the right lobe of the conductance curve (orange area in Fig. 6b): an increase in electrode displacement corresponds to a higher conductance (and a decrease to a lower conductance). Along a similar line, traces in antiphase with the gap modulation can be related to molecules which are precompressed in the starting configuration (blue panel in Fig. 6c and "Precompressed" video in SI) and therefore correspond to oscillations at the left lobe of the conductance curve (blue area in Fig. 6b). Traces with a doubled frequency (such as the green trace in Fig. 6a, and "Relaxed" video in SI) are related to molecules which are close to the relaxed gas phase geometry, in which the Fermi energy of the leads is aligned with the position of the interference dip (green panel, Fig. 6c). In this case the conductance dip is crossed two times during each piezo modulation period, therefore doubling the frequency of the measured conductance, as can be seen by following the purple steps in Fig. 6b. The appearance of the doubled frequency is thus a direct proof for the existence of the destructive interference dip. Importantly, the ability to mechanically tune the position of this dip to be located at the Fermi energy can be exploited in future studies and applications of quantum interference effects.

We note that the conductance variations as a function of displacement can be used to estimate the gauge factor, characterizing the piezoresistive response of the molecular spring. The gauge factor is defined as the relative conductance change divided by displacement normalized to the molecular length (see Fig. 3b and Supporting Section II-4 for more details). We find gauge factors exhibiting a wide distribution with a peak located at $GF = 5$ and a tail at higher values, reaching up to 40: orders of magnitude larger than those that have been reported on single DNA molecules.[37] We expect that thermally occupied ring rotations of the OPE rods at room temperature will reduce the electronic coupling in the molecule, the electron delocalization and hence the conductance. Besides known shortcomings of DFT with regard to the description of level alignments,[38] this could explain part of the overestimation of the theoretically predicted conductance values in our static DFT geometries. In addition, longitudinal vibrations, as well as thermal fluctuations, will lead to gap-size modulations and an effective averaging over a range of junction configurations. Such vibrationally induced decoherence effects will wash out the interference-induced conductance minimum.[39] Therefore a precise control of the temperature may turn out to be crucial in the optimization of gauge factors, and the molecule studied here appears to be an ideal candidate to investigate quantum interference effects at lower temperatures and to quantify if decoherence limits the room-temperature performance.[39–41] In order to achieve even higher gauge factors, it would also be interesting to explore different chemical designs based on the mechanical manipulation principle of quantum interference.

**Methods**

**MCBJ experiments**. For MCBJ experiments, a thin (<100 nm) gold wire is lithographically fabricated and suspended on a flexible substrate. Atomically sharp electrodes are formed when rupturing the wire in a three-point bending configuration. After breaking the wire, the electrodes can be fused together again by reducing the mechanical force used to bend the substrate. This opening and closing of the gap can be controlled with sub-nanometre accuracy on the position of the electrodes. Further details about the MCBJ setup and the measuring technique have been described elsewhere.[29,30] The molecule was dissolved in dichloromethane and the solution was drop casted on the MCBJ sample after the characterisation of the bare device (Fig. S10). All measurements were performed in air at room temperature after the evaporation of the solvent. Concentrations of 9 μM (sample **A**, **B**), 90 μM (sample **C**) and 900 μM (sample **D**) have been used, but no significant dependence on the concentration has been observed.

**Fast-breaking measurements.** Fast-breaking measurements were performed by applying a bias of 100 mV and using a constant pulling speed of the electrodes of 4.0 nm/s. The conductance is recorded until it falls below the noise level, which is about $5 \cdot 10^{-7}$ $G_0$ in our setup. At this point the electrodes are fused back and a new trace is recorded. 3,000 such traces were recorded for samples **A**, **B** and **C**; 5,000 for sample **D**. Further information on the technique can be found in Frisenda *et al.*[31]

**Distance-modulation measurements.** In these measurements, the electrode spacing was modulated to periodically increase and decrease.[32] The MCBJ was initially stretched to a few-atom width (3 $G_0$) and allowed to self-break by its surface tension to form atomically sharp electrodes. Then, the electrodes were separated by 1.75 nm, which is approximately the estimated length of the unstretched molecule. At this point, a 0.5 Hz triangular wave was applied to the piezoelectric stack that controls the electrode positions with a peak-to-peak gap size variation of 5.0 Å (Fig. 3) or 2.5 Å (Fig. S12), depending on the amplitude of the applied piezo voltage. Note, that a higher voltage on the piezoelectric corresponds to a larger electrode separation.

The initial opening of the junction allows to recognise if trapping of a molecule occurs, since for a displacement of 1.75 nm the tunnelling conductance in the absence of a molecule is below the noise level of our set-up. The modulation was kept for 120 s, after which the junction was stretched until the noise level was reached and fused back again to start a new cycle. It should be noted that the initial configuration in a distance-modulation experiment cannot be controlled.

**Simulations**. DFT calculations were carried out with TURBOMOLE.[42] We used the def-SVP basis set and the PBE functional.[43,44] During structural relaxations the total energy was converged to $10^{-6}$ a.u. and the maximum norm of the Cartesian gradient to $10^{-3}$ a.u. The electronic transmission was evaluated with a proprietary cluster-based framework. The procedure includes the separation of the system into semi-infinite leads and a central part, which contains the molecule and parts of the electrodes. The energy-dependent transmission function is expressed in terms of non-equilibrium Green's functions (NEGF) of the leads and the central part. The bulk parameters of the leads were extracted from a cluster of 1415 gold atoms. Further details on the method can be found in Pauly *et al.*[34]

## Acknowledgements

The authors in Basel and Delft acknowledge financial support by the European FP7-ITN MOLESCO (project no. 606728) and the H2020 FET QuIET (project no. 767187). The work at the University of Basel was also supported by the Swiss National Science Foundation (SNF, grant no. 200020-178808) and the Swiss Nanoscience Institute (SNI); the work at TU Delft by the EU through an advanced ERC grant (Mols@Mols). M.S. and F.P. acknowledge funding through the Collaborative Research Centre (SFB) 767 of the German Research Foundation (DFG) as well as computational resource provided by the state of Baden-Württemberg through bwHPC and the DFG through grant no INST 40/467-1 FUGG. Device fabrication was done at the Kavli Nanolab at Delft. D.S. thanks Riccardo Frisenda for fruitful discussions. K.J.W. thanks Markus Neuburger for single-crystal X-ray analysis.

## Author contributions

D.S., C.H., K.J.W., M.M. and H.S.J.Z. conceived and designed the experiment. M.L.P. assembled the MCBJ setup. M.L.P., D.S. and C.H. wrote the software used for the experiments. D.S. and C.H. performed the break junction experiments and the data analysis. K.J.W. and M.M. provided and characterised the molecule. M.S. and F.P. performed transport calculations. All authors wrote the manuscript.

**Supporting Information Available**

Details about the synthesis and characterisation of the molecule, additional fast-breaking and distance-modulation measurements, details about the statistical analysis, and on the transport calculations are available in the Supporting Information. This material is available free of charge via the Internet at http://pubs.acs.org.

**Competing interests**

The authors declare no competing financial interests.

**Materials & Correspondence**

Correspondence and request of materials regarding chemistry matters should be addressed to M.M.; those regarding theoretical calculations should be addressed to F.P. and those regarding the experiments to H.S.J.Z.

**TOC graphics**

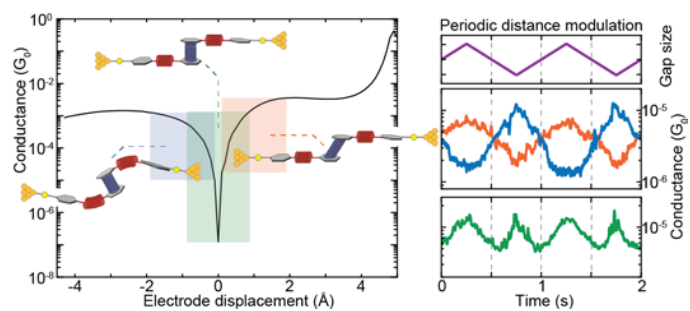